\documentclass[useAMS]{mn2e}

\newif\ifAMStwofonts
\AMStwofontstrue

\title[The Local and Accreted Mass Functions]{Super-massive Black Hole Demography: the Match between the Local and Accreted Mass Functions}

\author[F. Shankar
et al.]{F. Shankar$^1$, P. Salucci$^{1}$, G.L. Granato$^{1,2}$, G.
De Zotti$^{1,2}$, and
L. Danese$^{1}$\\
$^1$SISSA/ISAS, via Beirut 2, I-34014 Trieste, Italy\\
$^2$INAF - Osservatorio Astronomico di Padova, vicolo
dell'Osservatorio 5, I-35122 Padova, Italy\\}
\pagerange{\pageref{firstpage}--\pageref{lastpage}} \pubyear{2004}
\begin{document}
\maketitle \label{firstpage}

\newcommand{\lsim}{\mathrel{\rlap{\raise -.3ex\hbox{${\scriptstyle\sim}$}}%
                   \raise .6ex\hbox{${\scriptstyle <}$}}}%
\newcommand{\gsim}{\mathrel{\rlap{\raise -.3ex\hbox{${\scriptstyle\sim}$}}%
                   \raise .6ex\hbox{${\scriptstyle >}$}}}%

\begin{abstract}
We have performed a detailed analysis of the local super-massive
black-hole (SMBH) mass function based on both kinematic and
photometric data and derived an accurate analytical fit in the
range $10^6\leq M_{\rm BH}/M_{\odot}\leq 5\times 10^9$. We find a
total SMBH mass density of $(4.2\pm 1.1) \times 10^5
M_{\odot}/\hbox{Mpc}^3$, about 25\% of which is contributed by
SMBHs residing in bulges of late type galaxies.
%The local SMBH mass function in the range $10^6\leq M_{\rm
%BH}/M_{\odot}\leq 5\times 10^9$ is well represented by the
%expression $\Phi(M_{\rm BH}) d\log M_{\rm BH}=\Phi_{*} (M_{\rm
%BH}/M_{*})^{\alpha+1}\exp[-(M_{\rm BH}/M_{*})^{\beta}]$, with
%$\Phi_{*} = 7.7 (\pm 0.3) \cdot 10^{-3}$ Mpc$^{-3}$, $M_{*}=6.4
%(\pm 1.1) \cdot 10^7$ $M_{\odot}$, $\alpha = -1.11 (\pm 0.02)$ and
%$\beta = 0.49 (\pm 0.02)$ ($H_0 =
%70\,\hbox{km}\,\hbox{s}^{-1}\,\hbox{Mpc}^{-1}$).
Exploiting up-to-date luminosity functions of hard X-ray and
optically selected AGNs, we have studied the accretion history of
the SMBH population. If most of the accretion happens at constant
$\dot{M}_{\rm BH}/M_{\rm BH}$, as in the case of Eddington limited
accretion and consistent with recent observational estimates, the
local SMBH mass function is fully accounted for by mass accreted
by X-ray selected AGNs, with bolometric corrections indicated by
current observations and a standard mass-to-light conversion
efficiency $\epsilon \simeq 10\%$. The analysis of the accretion
history highlights that the most massive BHs (associated to bright
optical QSOs) accreted their mass faster and at higher redshifts
(typically at $z>1.5$), while the lower mass BHs responsible for
most of the hard X-ray background have mostly grown at $z<1.5$.
The accreted mass function matches the local SMBH mass function
if, during the main accretion phases, $\epsilon \simeq
0.09\,(+0.04,-0.03)$ and the Eddington ratio $\lambda=L/L_{\rm
Edd}\simeq 0.3\,(+0.3,-0.1)$ (68\% confidence errors). The
visibility time, during which AGNs are luminous enough to be
detected by the currently available X-ray surveys, ranges from
$\simeq 0.1\,$Gyr for present day BH masses $M_{\rm BH}^0\simeq
10^6\,M_{\odot}$ to $\simeq 0.3\,$Gyr for $M_{\rm BH}^0 \geq 10^9$
$M_{\odot}$. The mass accreted during luminous phases is $\geq
25$--$30\%$ even if we assume extreme values of $\epsilon$
($\epsilon \simeq 0.3-0.4$). An unlikely fine tuning of the
parameters would be required to account for the local SMBH mass
function accomodating a dominant contribution from 'dark' BH
growth (due, e.g., to BH coalescence).
\end{abstract}


\section{INTRODUCTION}

The paradigm that quasars and, more generally, Active Galactic
Nuclei (AGNs) are powered by mass accretion onto a super-massive
black hole (SMBH) proposed long ago (Salpeter 1964; Zeldovich \&
Novikov 1969; Lynden--Bell 1969) has got very strong support from
spectroscopic and photometric studies of the stellar and gas
dynamics in the very central regions of local spheroidal galaxies
and prominent bulges. These studies established  that in most, if
not all, galaxies observed with high enough sensitivity a central
massive dark object (MDO) is present with a well defined
relationship between the MDO mass and the mass or the velocity
dispersion of the host galaxy spheroidal component (Kormendy \&
Richstone 1995; Magorrian et al. 1998; Gebhardt et al. 2000;
Ferrarese \& Merritt 2000; Tremaine et al. 2002; Kormendy 2003).
Although there is no direct evidence that all MDOs are black holes
(BHs), the evidence for a singularity is actually very tight in
the Galaxy (Sch\"odel et al. 2002; Ghez et al. 2003) and
alternative explanations are severely constrained in NGC 4258
(Miyoshi et al. 1995; see also e.g. Kormendy 2003).

Soltan (1982) showed that, in the framework of the above paradigm,
the total accreted mass density can be inferred from the observed
QSO/AGN counts. The basic ingredients of the calculation are {\it
i)} the bolometric correction $k_{\rm bol}$, {\it ii)} the
effective redshift and the corresponding K-correction, and {\it
iii)} the mass to radiation conversion efficiency $\epsilon$. When
more precise luminosity functions of QSO/AGN became available,
Chokshi \& Turner (1992) presented a first estimate of the
accreted mass density and derived constraints on the corresponding
mass function (MF). Small \& Blandford (1992) tried to relate the
luminosity function (LF) of optically selected AGNs to the local
SMBH MF in galaxies for several possible accretion histories.
Salucci et al. (1999) showed that the LF of the galaxy spheroids
(encompassing E and S0 galaxies, and bulges of spiral galaxies)
combined with the relationship between the spheroid and the
central MDO mass allows an accurate evaluation of the local MF of
the SMBHs. They concluded that the distribution of the mass that
fuelled the nuclear activity, as traced by the LFs of QSOs and of
AGNs contributing the X-ray background, matches the local SMBH MF,
provided that $\epsilon \simeq 0.1$ and the ratio of bolometric to
Eddington luminosity $\lambda = L_{\rm bol}/L_{\rm Edd}$ declines
with luminosity and/or look-back time. It was also shown that the
high mass tail of the local MF is due to the remnants of the
bright optically selected QSOs, while at low masses the remnants
of the AGNs producing most of the X-ray background largely
dominate.

Recent estimates of the local MF also exploited the relationship
between SMBH mass and velocity dispersion of the host spheroid
(see e.g. Yu \& Tremaine 2002; Aller \& Richstone 2002; McLure \&
Dunlop 2003; Marconi et al. 2004), to reach, however, somewhat
discrepant conclusions.  %McLure \& Dunlop (2003) presented virial
%SMBH mass estimates for more than 12000 QSOs in the redshift
%interval $0.1\leq z\leq 2.1$, with ensuing evaluation of the
%$L/L_{Edd}$ ratios.

The aim of this paper is to compare the local MF of SMBHs with the
MF of the mass accreted during the nuclear activity (AMF), in
order to shed light on important open questions such as:
\begin{itemize}
\item  is there room for a significant 'dark'
accretion\footnote{By 'dark' accretion we mean accretion not
traced by either optical or X-ray surveys. This is quite different
from the often used term of ``obscured'' accretion, which is
referred to accretion on type 2 AGNs (see e.g. Fabian 2003).}? In
particular, can BH merging be the main process building the
present day SMBH MF?

\item  how do the X-ray and optical views of BH mass build-up
compare?

\item  what is the typical Eddington ratio of AGNs when they
accrete most of their mass?

\item how long does the visible phase of the AGNs last?

 \item do we really need a mass-to-radiation conversion efficiency higher than
the standard value of $\epsilon \simeq 0.1$, as recently argued
(see e.g. Elvis, Risaliti \& Zamorani 2002)?

\end{itemize}

The local SMBH MF, including the contribution from the spheroidal
components of late-type galaxies, is re-estimated exploiting and
extending the technique outlined by Salucci et al. (1999) and
recently presented by Shankar et al. (2003). The accreted MF will
be derived using up-to-date LFs as a function of cosmic time for
optically and X-ray selected AGNs.

The paper is organized as follows. In  Section 2 we critically
discuss the relationships among luminosity, mass, and velocity
dispersion of the spheroidal components of galaxies ($L_{\rm
sph}$, $M_{\rm sph}$, and $\sigma$), and $M_{\rm BH}$.  In Section
3 we present and discuss two estimates (derived via the velocity
dispersion distribution function (VDF) and the LF, respectively)
of the local SMBH MF. In Section 4 the accreted mass density is
estimated. In Section 5 the accreted mass function (AMF) is
computed and compared with the local MF of SMBHs. The accretion
history and the AGN visibility times are analyzed in Section 6. A
discussion of the results and the conclusions are presented in
Section 7.

The standard flat $\Lambda CDM$ cosmological model has been used,
with $H_0 = 70\,\hbox{km}\,\hbox{s}^{-1}\,\hbox{Mpc}^{-1}$,
$\Omega_{m}=0.3$, $\Omega_{\Lambda}=0.7$.

\section{Correlations among SMBH mass, galaxy luminosity and velocity dispersion}
\label{sect2}

The SMBHs MF can be derived coupling the statistical information
on local LFs of galaxies with relationships among luminosity (or
related quantities, such as stellar mass and velocity dispersion)
and the central BH mass (see e.g. Salucci et al. 1999).

Since the BH mass correlates with luminosity and velocity
dispersion of the bulge stellar population, we need separate LFs
for different morphological types (which have different bulge to
total luminosity ratios), and it is convenient to use galaxy LFs
derived in bands {\it as red as possible}, where the old bulge
stellar populations are more prominent.

\subsection{Bulge luminosity versus black hole mass}

McLure \& Dunlop (2002) analyzing a sample of 72 active and  20
inactive galaxies found that the central BH mass and the total
R-band magnitude, $M_R$, of the bulge are strictly related. In
particular, considering only inactive elliptical galaxies with
accurate measurements of BH mass, the relation, converted to
$H_0=70$, reads
\begin{equation} \log(\frac{M_{\rm BH}}{M_{\odot}})=-0.50(\pm
0.05)M_{R}-2.69(\pm 1.04) \ , \label{DML}
\end{equation}
with a scatter of $\Delta \log(M_{\rm BH})=0.33$. It is worth
noticing that the relationship has been derived using B-band
magnitudes, translated to R-band
%computed in B-band and then translated to R-band
assuming an average color (B-R)=1.57. A larger scatter $\Delta
\log(M_{\rm BH}) \simeq 0.45$ was found by Kormendy \& Gebhardt
(2001), who used a sample including also lenticular and spiral
galaxies.

For galaxies observed with a spatial resolution high enough to
resolve the BH sphere of influence,  Marconi \& Hunt (2003) report
a tight relation between the SMBH mass and the host galaxy bulge
K-band luminosity
\begin{equation}
\log(\frac{M_{\rm BH}}{M_{\odot}})=1.13(\pm 0.12)
[\log(\frac{L_{K}}{L_{K_\odot}})-10.9]+8.21(\pm 0.07) \label{MBHK}
\end{equation}
with a scatter $\Delta \log M_{\rm BH} = 0.31$. Translating
Eq.~(\ref{DML}) to the K-band using the colour $R-K=2.6$ (Kochanek
et al. 2001, with $K-K_{20}=-0.2$) as discussed in Sect. 3.1, it
is apparent that the Marconi \& Hunt (2003) relationship yields
higher BH masses at fixed luminosity. Correspondingly, the derived
SMBH mass density is up to a factor of 2 higher than obtained from
Eq.~(\ref{DML}). A closer analysis shows that most of the
discrepancy is due to SMBH in spiral galaxies and can be ascribed
to the uncertainty in the evaluation of their bulge component.
Since most of the local mass density is contributed by BHs in E
and S0 galaxies, we decided to exploit the relationship reported
in Eq.~(\ref{DML}).
%The difference amounts to a
%factor of about 3 at $L_K/L_{K_\odot}=10^9$, and decreases at
%higher luminosities, becoming a factor of about 1.6 at
%$L_K/L_{K_\odot}=10^{12}$. However, if we consider  only  E and S0
%galaxies, their relationships $M_{BH}-L_B$ and $M_{BH}-L_K$ ,
%translated to R-band, agree with Eq.~(\ref{DML}) within 20$\%$.

As pointed out by McLure \& Dunlop (2002), their $M_R-\log M_{\rm
BH}$ relation is compatible with a linear relation between BH and
spheroidal mass, $M_{\rm sph}$. Indeed, inserting the result found
by Borriello et al. (2003), $M_{\rm sph}/L_R \propto L_R^{0.21 \pm
0.03}$, we get $M_{\rm BH}\propto M_{\rm sph}^{1.03\pm 0.12}$.

\subsection{Velocity dispersion versus BH mass}

While the presence of a strong correlation between BH mass and
velocity dispersion of the stellar spheroid, $M_{\rm BH}-\sigma$,
is undisputed (Ferrarese \& Merritt 2000; Gebhardt et al. 2000),
the value of its slope is still debated. A detailed analysis of
the available data by Tremaine et al. (2002) yields
\begin{equation}
\log(M_{\rm BH}\frac{80}{H_{0}})=4.02 (\pm 0.32)
\log(\sigma_{200})+8.13(\pm 0.06) \ , \label{Tremaine}
\end{equation}
$\sigma_{200}$ being the line-of-sight velocity dispersion in
units of $200\,\hbox{km}\,\hbox{s}^{-1}$. The slope is in
reasonable agreement with the findings of Ferrarese (2002),
$M_{\rm BH} \propto \sigma^{4.58 \pm 0.52}$. It should also be
noted that the velocity dispersions used by Tremaine et al. (2002)
refer to a slit aperture $2r_e$, while those reported by Ferrarese
refer to $r_e/8$. The scatter around the mean relationship is
small, $\Delta \log M_{\rm BH}=0.3$, possibly consistent with pure
measurement errors.

The low mass and low velocity dispersion regime is quite difficult
to investigate. The analysis of M33 by Gebhardt et al. (2001)
yields an upper limit on the BH mass ($\sim 1500 M_{\odot}$) more
than 10 times below the central value predicted by
Eq.~(\ref{Tremaine}). However the larger upper limit ($\sim 3000
M_{\odot}$) claimed by Merritt, Ferrarese \& Joseph (2001) is
consistent with the steeper $M_{BH}-\sigma$ relation found by
Ferrarese (2002). Filippenko \& Ho's (2003) estimate of the mass
of the central BH in the least luminous type 1 Seyfert galaxy
known, NGC 4395, ($M_{\rm BH}\simeq 10^4$-- $10^5\,M_{\odot}$) is
not inconsistent with Eq.~(\ref{Tremaine}). However it should be
also mentioned that the BH mass in this case has been estimated
using indirect, rather than dynamical arguments. The efforts to
detect the so called intermediate mass BHs ($10^3\,M_{\odot} \lsim
M_{\rm BH}\lsim 10^6\,M_{\odot}$) in galactic centers and
therefore to constrain the very low $\sigma$ ($<
50\,\hbox{km}\,\hbox{s}^{-1}$) end of the $M_{\rm BH}$--$\sigma$
relation have been recently reviewed by van der Marel (2003).

It is worth noticing that the $M_{\rm BH}$--$\sigma$ relationship
needs not to keep a power-law shape down to low $M_{\rm BH}$ or
$\sigma$ values. On the contrary, Granato et al. (2004) presented
a model for the coevolution of QSOs and their spheroidal hosts
whereby, in the least massive bulges, the BH growth is
increasingly slowed down by supernova heating of the interstellar
medium as the bulge mass (hence $\sigma$) decreases. As a result,
$M_{\rm BH}$ is expected to fall steeply with decreasing $\sigma$,
for $\log[\sigma(\hbox{km}\,\hbox{s}^{-1})]\lsim 2.1$. This model
also predicts that the observed spread around the mean
relationship is a natural one, deriving mainly from the different
virialization redshifts of host halos.

\begin{figure}
\begin{center}
\includegraphics{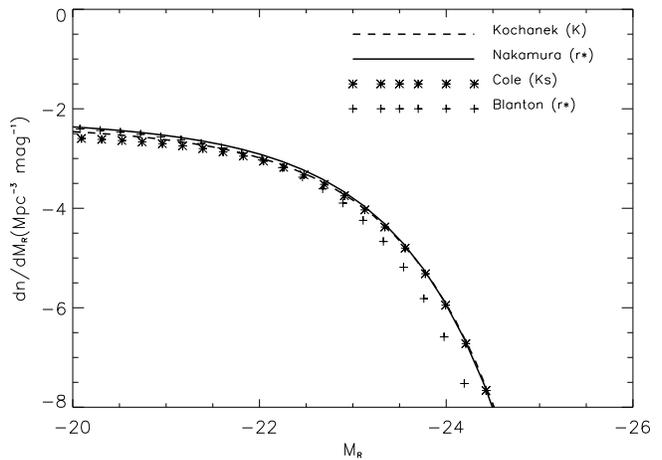} \vspace{6.4cm} \caption{Galaxy luminosity function
estimates converted to R-band total magnitudes as described in the
text.} \label{fig:LFRtot}
\end{center}
\end{figure}

\begin{figure}
\begin{center}
\includegraphics{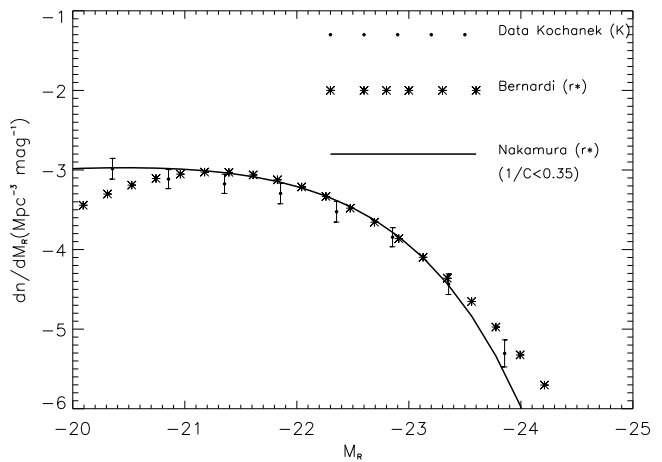} \vspace{6.4cm} \caption{R-band luminosity function
estimates  for early-type galaxies. Data points from Kochanek et
al. (2001).} \label{fig:LFearly}
\end{center}
\end{figure}

\begin{figure}
\begin{center}
\includegraphics{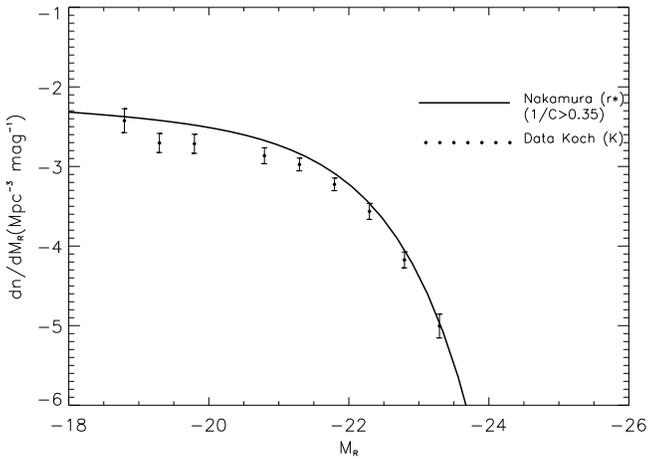} \vspace{6.4cm} \caption{R-band local luminosity
functions estimates for late-type galaxies.} \label{fig:LFlate}
\end{center}
\end{figure}

\section{The local SMBH  Mass Function}
\label{SMBHMF}

The local SMBH MF can be estimated either from the local LF or
from the local velocity dispersion function (VDF) of spheroidal
galaxies and galaxy bulges, through the $M_{\rm BH}$--$L_{\rm
sph}$ or the $M_{\rm BH}$--$\sigma$ relation, respectively.
Previous studies (Yu \& Tremaine 2002; Aller \& Richstone 2002)
have shown that the two methods may yield estimates of the local
mass density of SMBHs differing by a factor of $\simeq 2$. On the
other hand Ferrarese (2002), McLure \& Dunlop (2003) and Marconi
et al. (2004) found very good agreement among the results of the
two methods.

\subsection{Local luminosity functions of spheroids and bulges}
\label{LLF}

The LFs best suited for our purpose are those in red and IR bands,
which are more directly linked to the mass in old stars. Moreover,
we need separate LFs for the various morphological types with
different bulge to total luminosity ratios. We will use the K-band
LF by Kochanek et al. (2001), the $K_{s}$-band LF by Cole et al.
(2001), the $r^*$ band LF by Blanton et al. (2001, 2003), Nakamura
et al. (2002), and Bernardi et al. (2003). To compare LFs defined
in different bands we must set up a common definition of the total
magnitude/luminosity and of average colours.

Since we are interested in the total luminosity of spheroidal
components of galaxies, we adopt as total magnitudes those
obtained with a de Vaucouleurs profile. We have therefore
corrected by $-0.2$ the surface brightness limited magnitudes,
$K_{20}$, of the 2MASS sample and the Petrosian magnitudes, used
by Blanton et al. (2001; 2003) and by Nakamura et al. (2002). Both
magnitude systems in fact are defined for apertures which contain
$\sim 80\%$ of the total flux for the adopted profile. For the
Kron magnitudes of Cole et al. (2001), in the $K_s$ band, we used
a brightening of $-0.11$, which is required to convert them to an
$r^{1/4}$ luminosity profile.

Magnitudes were converted to the R-band using the mean colours
$R-K_s=2.51$ and $R-r^\star=-0.11$ (Blanton et al. 2001). We
assume an error of 0.1 mag. on colours and include it in our
estimate of the final errors on the SMBH MF.

As illustrated by Fig.~\ref{fig:LFRtot}, the different estimates
of the LF are in very good agreement with each other, except for
that by Blanton et al. (2003), which is low at bright magnitudes
(by a factor $\simeq 4$ at $M_R < -24$). Indeed the Schechter
function adopted by the latter authors falls below their own data
points at $M_{r^*}-5\log(H_0/100)= -23$ (cfr. their Fig. 5).  The
classification by Kochanek et al. (2001) allows a clear cut
distinction between early and late type galaxies. A similar
classification has also been proposed by Nakamura et al. (2002).
Figures~\ref{fig:LFearly} and \ref{fig:LFlate} show that the
agreement is quite satisfactory also for early and late types
separately, although the early-type LF by Bernardi et al. (2003)
misses objects fainter than $M_R \simeq -21$ because of their
velocity dispersion criterion ($\sigma
>70\,\hbox{km}\,\hbox{s}^{-1}$ and $S/N > 10$).

\begin{figure}
\begin{center}
\includegraphics{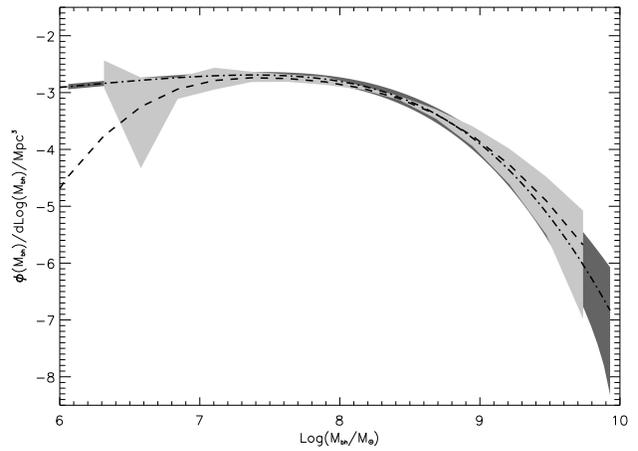} \vspace{6.4cm} \caption{Local mass function of SMBHs
hosted by early-type galaxies. The dot-dashed line shows the
estimate obtained from the $r^*$-band LF (Nakamura et al. 2002)
coupled with the $M_{\rm BH}$--$L_{\rm bulge}$ relation (see
text); the dark gray area shows the estimated errors. The dashed
line and light gray area refer the MF derived using the bivariate
dispersion velocity distribution (cfr.
Fig.~\protect\ref{SMBHMF2}).} \label{SMBHMF1}
\end{center}
\end{figure}

\begin{figure}
\begin{center}
\includegraphics{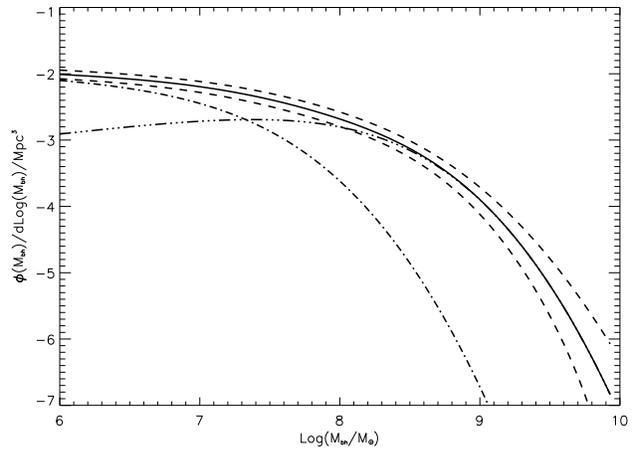} \vspace{6.4cm} \caption{Global SMBH mass function
(solid line) and its uncertainties (dashed lines). The three
dots-dashed and the dot-dashed lines show the contributions from
SMBHs hosted by early-  and late-type galaxies, respectively. }
\label{SMBHMFtot}
\end{center}
\end{figure}

\subsection{From the local luminosity function to the SMBH mass
function}

Based on Table~1 of Fukugita, Hogan \& Peebles (1998), to obtain
the LF of the spheroidal components we adopt the average $R$-band
bulge-to-total luminosity ratios $f_{\rm sph}^{\rm early}=0.85 \pm
0.05$ for early-type galaxies and $f_{\rm sph}^{\rm late}=0.30 \pm
0.05$ for late-type galaxies.

The SMBH mass function was then computed convolving the LF with
the $M_{\rm BH}$--$L_{\rm  sph}$ relation by McLure \& Dunlop
(2002) [see Eq.~(\ref{DML})], assuming a Gaussian distribution
around the mean with a dispersion $\Delta \log(M_{\rm
BH})=0.33^{+0.07}_{-0.05}$. These uncertainties encompass most of
the values quoted in the recent literature (see Sect.~2.2). The
errors on the SMBH MF include the overall uncertainties on the LF,
on the bulge fractions, on the $M_{\rm BH}$--$L_{\rm  sph}$
relation and its scatter. The uncertainties on the LF, including
those on colours, contribute about $70\%$ of the error budget.

The SMBH MF in early type galaxies, shown in Fig.~\ref{SMBHMF1},
has been estimated using the LF of Nakamura et al. (2002),
converted to R-band, and Eq.~(\ref{DML}). The match with the MF
computed via the VDF and the $M_{\rm BH}$--$\sigma$ relation (see
below) is very good. The corresponding SMBH mass density amounts
to $\rho _{\rm BH}^{0}(E)=3.1 ^{+0.9}_{-0.8} \times 10^5
M_{\odot}/\hbox{Mpc}^3$, in excellent agreement with the findings
of McLure \& Dunlop (2003) and Marconi et al. (2004), and 30$\%$
higher than the estimate by Yu \& Tremaine (2002) and Aller \&
Richstone (2002).

The MF of SMBH hosted by spiral bulges was computed in the same
way, using the LF for late-type galaxies by Nakamura et al.
(2002). Their local mass density is $\rho _{\rm BH}^{0}(Sp)=(1.1
\pm 0.5) \times 10^5 M_{\odot}/\hbox{Mpc}^3$, bringing the overall
mass density to $\rho _{\rm BH}^{0}=(4.2\pm 1.1) \times 10^5
M_{\odot}/\hbox{Mpc}^3$. The local number density of SMBHs with
$M_{\rm BH}>10^7\, M_\odot$ is $n_{\rm SMBH}\simeq (1.3 \pm 0.25)
\times 10^{-2}$ $\hbox{Mpc}^{-3}$. As illustrated by
Fig.~\ref{SMBHMFtot} the main contribution to the global mass
density comes from the range $2\times 10^{7} < M_{\rm BH} <
1\times 10^9 M_{\odot}$, mostly populated by SMBH in early-type
galaxies, while less massive BHs are preferentially hosted in late
type objects.

Our determination is very close to the result by Marconi et al.
(2004), who used a methodology  similar to ours. As suggested by
Aller \& Richstone (2002) the MF can be well represented by a four
parameter function, which for our determination (per unit $d\log
M_{\rm BH}$) takes the form:
\begin{equation}
\Phi(M_{\rm BH})= \Phi_{*}\left(\frac{M_{\rm
BH}}{M_{*}}\right)^{\alpha+1} \exp\left[-\left(\frac{M_{\rm
BH}}{M_{*}}\right)^{\beta}\right], \label{MFfit}
\end{equation}
\noindent with $\Phi_{*} = 7.7 (\pm 0.3) \cdot 10^{-3}$
Mpc$^{-3}$, $M_{*}=6.4 (\pm 1.1) \cdot 10^7$ $M_{\odot}$, $\alpha
= -1.11 (\pm 0.02)$ and $\beta = 0.49 (\pm 0.02)$ ($H_0 =
70\,\hbox{km}\,\hbox{s}^{-1}\,\hbox{Mpc}^{-1}$). The formula holds
in the range $10^6 \leq M_{\rm BH}/M_{\odot} \leq 5\times 10^9$.

\begin{table}
\caption{Local SMBH mass densities} \centering
 \begin{tabular}{lc}
 \hline
Method   & $\rho^0_{\rm BH}(10^{5} M_{\odot}\hbox{\rm Mpc}^{-3}
h_{70}^2 )$
                                         \\ \hline
 %\bfseries{Paper I  (early)}        &                         &                                 &   $5.2$               \\ \hline
% \bfseries{Kochanek (K)}        &        $3.3  $                                   \\ \hline
                              \bfseries{Early Type Galaxies}   & \\
%\hline
$r^{*}$+$M_{\rm BH}-L_{\rm  bulge}$&    $3.1^{+0.9}_{-0.8}$        \\
%\hline
 %\bfseries{Nakamura (r*band)}       &        $2.6  $                     \\ \hline
$\hbox{bivariate~VDF} + (M_{\rm BH}\hbox{--}\sigma)$ & $3.0^{+1.0}_{-0.6}$     \\
%\hline
$\hbox{Sheth~VDF} + (M_{\rm BH}\hbox{--}\sigma)$            & $2.8$     \\
%\hline \bfseries{Gonzalez VDF +$M_{\rm BH}-\sigma$}        &
%$3.7 $
\\ \hline                   \bfseries{Late Type Galaxies}        &
%\\ \hline \bfseries{late type $r^{*}$+$M_{\rm BH}-\sigma$}        &        $0.9-1.66 $
\\%\hline
$r^{*}+ (M_{\rm BH}\hbox{--}L_{\rm  bulge})$        & $1.1 \pm 0.5
$
\\ \hline

%\bfseries{late type (Kband)}        &        $0.5 $      \\\hline
\end{tabular} \label{rhoBH}
\end{table}

\begin{figure}
\begin{center}
\includegraphics{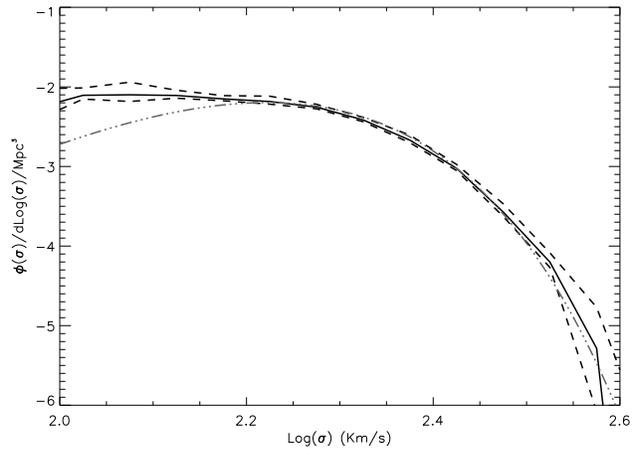} \vspace{6.4cm} \caption{Velocity dispersion
function. The solid line is the estimate obtained from the
Nakamura et al. (2002) LF coupled with the bivariate (luminosity,
$\sigma$) distribution derived from the SDSS data in the
$r^{*}$-band; its uncertainty region is shown by the dashed lines.
The three dots-dashed line is the estimate by Sheth et al.
(2003).} \label{fig:vdf}
\end{center}
\end{figure}

\begin{figure}
\begin{center}
\includegraphics{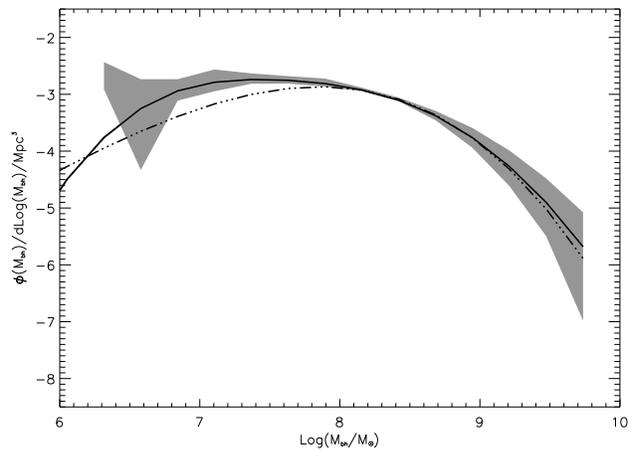} \vspace{6.4cm} \caption{Estimates of the local mass
function of SMBH hosted by early type galaxies, derived from the
velocity dispersion functions in Fig.~\protect\ref{fig:vdf}
coupled with the $M_{\rm BH}$--$\sigma$ relation by Tremaine et
al. (2002). The gray area represents the uncertainties on the
estimate based on the bivariate (luminosity, $\sigma$)
distribution.} \label{SMBHMF2}
\end{center}
\end{figure}

%\begin{figure}
%\begin{center}
%\special{psfile=./Radio_LF.ps hscale=40 vscale=40 hoffset=10
%voffset=-280} \vspace{5cm} \caption{data: stars Magliocchetti et al. (2002); crosses (Sadler (2003); fit: Dunlop \&
%Peacock (1990)}
%\end{center}
%\end{figure}

%\begin{figure}
%\begin{center}
%\special{psfile=./Radio_SMBH_MF.ps hscale=40 vscale=40 hoffset=10
%voffset=-280} \vspace{5cm} \caption{gray area: local SMBH MF and
%its uncertainty; points: SMBH from Sadler fit}
%\end{center}
%\end{figure}

%\begin{figure}
%\begin{center}
%\special{psfile=./FBH_Late_Comp.ps hscale=40 vscale=40 hoffset=10
%voffset=-280} \vspace{5cm} \caption{Contribution from late type
%SMBH. Dot-dashed line: Late type SMBH mass function obtained from
%the $M_{\rm BH}-L_{\rm  bulge}$ relation with its error bars (dashed
%lines); solid line: SMBH mass function obtained from the
%Giovanelli et al. (1997) $L-V_C$ fit and the Tremaine et al.
%(2002) $M_{\rm BH}-\sigma$ relation using Ferrarese (2002) $\sigma
%=0.61 V_C$}
%\end{center}
%\end{figure}

\subsection{The local velocity dispersion function} \label{sect3}

The local VDF can be derived from the local galaxy LF exploiting
the luminosity--$\sigma$ relation (Gonzalez et al. 2000; Sheth et
al. 2003), well established for spheroidal galaxies (Faber \&
Jackson 1976). The analysis of a sample of 86 nearby E and S0
galaxies, yields (de Vaucouleurs \& Olson 1982; Gonzalez et al.
2000):
\begin{equation}
 M_{B_{T}}=(-19.71 \pm 0.08)-(7.7 \pm 0.7)\log \sigma_{200}+5 \log
 h , \label{MBsigma}
\end{equation}
with $h=H_0/100\,\hbox{km}\,\hbox{s}^{-1}\,\hbox{Mpc}^{-1}$.
However, data for larger samples suggest a steeper relation.
Bernardi et al. (2003), using about 9000 early type galaxies
selected from the  Sloan Digital Sky Survey (SDSS), found $L_{r^*}
\propto \sigma^{3.91}$, where $\sigma$ refers to a $r_e/8$
aperture. The VDF of the SDSS has been actually obtained with a
fixed aperture of $1.^{"}5$ and then converted to the $r_e/8$
aperture following the conversion suggested by J\o rgensen, Franx
\& Kjaergaard (1995).

Estimates of the local VDF have been derived by Shimasaku (1993)
and Gonzalez et al. (2000) (see Kochanek 2001 for a comprehensive
review), and  more recently, by Sheth et al. (2003) who were the
first to allow for the distribution (assumed Gaussian with a
luminosity dependent width) of data points around the best fit
relationship.

To make a fuller exploitation of the data by Bernardi et al.
(2003) we have used them to derive the bivariate distribution
$p_{ij}=p(L_i,\sigma_j)$, yielding the fraction of objects in the
$r^*$-luminosity bin centered at $L_{i}$ and in velocity
dispersion bin centered at $\sigma_j$. The 9000 objects in the
samples, covering an absolute magnitude range $-18 \le M_{r^*} \le
-27$ and a velocity dispersion range $1.8 \le \log(\sigma) \le
2.7$ ($\sigma$ in km/s), have been subdivided in bins of width
0.05 both in $M_{r^*}$ (19 bins) and in $\log(\sigma)$ (170 bins).
The VDF is then estimated as:
\begin{equation}
n(\sigma_{j})=\sum_{i} p_{ij}n_{i} , \label{VDF}
\end{equation}
$n_i=n(L_i)$ being the $r^*$-band LF for early type galaxies by
Nakamura et al. (2002; Fig.~\ref{fig:LFearly}). The resulting VDF
is shown in Fig.~\ref{fig:vdf} with its errors, computed using the
formula for the propagation of errors in a multivariate function
with independent random errors in each variable. The uncertainties
are bigger towards the two extremes, where the number of sampled
objects decreases, and smaller around the knee of the function. We
have checked that our results are independent of the bin size.

Using the $K$-band LF (Kochanek et al. 2001) converted to the
$r^*$-band adopting a colour $K-r^{*}= -2.73$, appropriate for
early-type galaxies (Blanton et al. 2001; Kochanek et al. 2001) we
find differences in the VDF of at most 0.15 dex. From
Fig.~\ref{fig:vdf} it is apparent that our estimate is very close
to that by Sheth et al. (2003), apart for the more rapid decline
at low velocity dispersions, due to the selection criteria adopted
by Bernardi et al. (2003), as noted above.

% Moreover we have gone trough an
%analysis even including redshift, as the SDSS data are in the
%range $0<z<0.3$. We have redone the above calculations cutting off
%those objects with redshift greater than a certain value: we
%noticed the trend in the DVF to shift the knee towards lower
%$\sigma$, which is an expected effect due to the lower
%contribution of high $L/\sigma$ objects (E and big S0 galaxies) at
%lower redshift.

The contribution from late type galaxy bulges to the VDF is rather
difficult to assess. In fact, the bulge-to-disk mass ratios depend
more on morphology than on luminosity and on rotational velocity.
Although for a given morphological type a correlation between the
bulge velocity dispersion and the maximum rotational velocity may
be expected, the use of the Tully-Fisher relation (among
luminosity and rotation velocity) to infer the velocity dispersion
is rather unsafe (see discussions of Sheth et al. 2003 and
Ferrarese 2002).

\subsection{From the VDF to the SMBH MF}

In order to get an estimate of the local mass function of SMBHs,
the local VDF for early-type galaxies can be convolved with the
$M_{\rm BH}$--$\sigma$ relation of Tremaine et al. (2002). The
SDSS velocity dispersions (Bernardi et al. 2003) given for an
aperture of $r_e/8$, have been converted to $2r_e$ aperture using
Eq. (16) of Tremaine et al. (2002), while the $M_{BH}-\sigma$
relation of Tremaine et al (2002) has been estimated with velocity
dispersions taken within an aperture corresponding to 2$r_e$. We
assume a Gaussian distribution of BH masses at constant $\sigma$,
with a dispersion $\Delta= 0.30^{+0.07}_{-0.03}$ dex.

The SMBH MF estimates derived from the VDF obtained through the
bivariate probability distribution  and from the VDF by Sheth et
al. (2003) are shown in Fig.~\ref{SMBHMF2}. The shaded area shows
the uncertainties on the former estimate, including the
contributions from errors on both the VDF and $\Delta$. Again, the
decline for $M_{\rm BH}\leq 10^7\, M_{\odot}$ is due to the
incompleteness of the SDSS sample at low velocity dispersions. The
integrated mass density of SMBH in early-type galaxies is $\rho
_{\rm BH}^{0}=2.8 \times 10^5 M_{\odot}/\hbox{Mpc}^3$ or $\rho
_{\rm BH}^{0}= 3.0^{+1.0}_{-0.6} \times 10^5
M_{\odot}/\hbox{Mpc}^3$ if we use the VDF by Sheth et al. (2003)
or that obtained through the bivariate probability function,
respectively. As noted above, the evaluation of the contribution
of SMBHs hosted by late-type galaxies through this method is
hampered by the poor knowledge of the local VDF for their bulges.
Adopting the temptative estimate by Sheth et al. (2003) for the
late type galaxy contribution to the VDF, coupled with
Eq.~(\ref{Tremaine}), with the same scatter $\Delta=0.3$, we get
$\rho^0_{BH}=1.2\times 10^5$ $M_{\odot}/\hbox{Mpc}^3$, nicely
consistent with the estimate derived from the LF.

Wyithe \& Loeb (2003) obtained a lower estimate of the total mass
density mainly because they neglected the scatter $\Delta$ of the
$M_{BH}-\sigma$ relationship.

\section{The accreted mass density}

\indent Soltan (1982) showed that the total mass density
accumulated by accretion on BHs powering QSOs can be deduced from
QSO counts, under quite simple assumptions. If $\epsilon$ is the
mass to radiation conversion efficiency, the bolometric luminosity
is
\begin{equation}
L_{\rm bol}=\epsilon \dot{M}_{\rm acc}c^2 \label{Lbol}
\end{equation}
and the mass accretion rate reads
\begin{equation}
\dot{M}_{\rm BH}=(1-\epsilon)\dot{M}_{\rm acc}.
\end{equation}
The conversion of luminosities measured in a given band to
bolometric luminosities requires the knowledge of bolometric
corrections $k_{\rm bol}$ (see, e.g., Elvis et al 1994), which may
depend on luminosity and/or redshift. The mass accreted up to the
present time by all AGNs brighter than $L$ can be written as
\begin{equation}
\rho_{\rm acc}(>L) = \frac{1-\epsilon}{\epsilon c^2}\!\!
\int_{0}^{z_{\rm max}}\!\!\!\!\!\!dz\frac{dt}{dz} \int_{L}^{L_{\rm
max}}\!\!\!\!\!\!\!\!\!\!\!\!dL' k_{\rm bol}(L',z) n(L',z)L' .
\label{rhoac}
\end{equation}
where $n(L',z)$ is the {\it comoving} luminosity function. As
noted by Soltan (1982), $\rho_{\rm acc}$ is independent of $H_{0}$
and of the QSO lifetime.

The most complete AGN surveys are those at X-ray (hard and soft),
optical, and radio wavelengths. The latter selection is however
rather inefficient, since only $\sim 10\%$ of AGNs are radio loud.

\begin{figure}
\begin{center}
\includegraphics{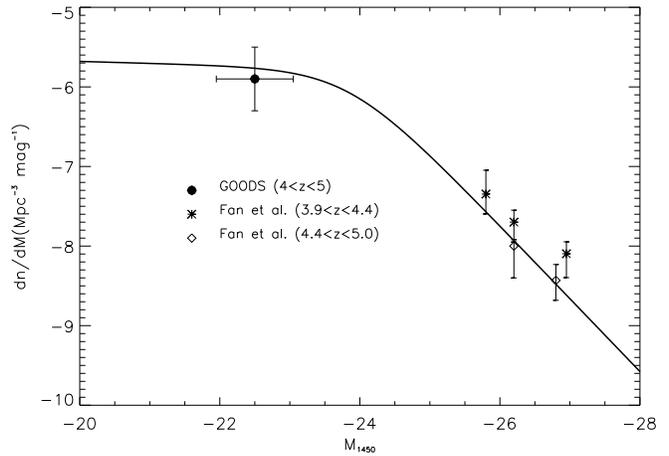} \vspace{6.4cm} \caption{Optical AGN LF at high-$z$.
Data from the SDSS (Fan et al. 2001) and GOODS survey (Cristiani
et al. 2004). The solid line shows the Croom et al. (2003)
power-law model at $z=4.5$.} \label{fig:opticalLF}
\end{center}
\end{figure}

\subsection{Mass accreted on optically selected QSOs}

The 2dF survey (Boyle 2000; Croom et al. 2003) has provided an
accurate determination of the redshift-dependent LF of optically
selected AGNs with $M_B<-22.5$ and $z<2.2$. Croom et al. (2003)
showed that the data are consistent with pure luminosity evolution
of the form $L_B(z)=L_B(0)\times 10^{0.21z(5.476-z)}$ (for a
$\Lambda$CDM model with $\Omega_m=0.3$), peaking at $z_p\simeq
2.74$ and exponentially declining at higher redshifts. Although
the luminosity function is poorly known for $z>2.4$, there is
strong evidence (see Fan et al. 2001 and Osmer 2003 for a recent
review), for a rapid decrease with increasing redshift of the
space density of bright QSO for $z \ge 3$. More recently, very
deep X-ray (Barger et al. 2003) and optical (Cristiani et al.
2004) surveys have provided strong constraints on the space
density of less luminous QSOs at high redshift. As illustrated by
Fig.~\ref{fig:opticalLF}, the Croom et al. (2003) power-law model
provides a sufficiently accurate description also of the data at
$z \ge 4$.

%Although higher redshift data (Fan et al. 2001) suggest deviations
%from pure luminosity evolution, the Croom et al. (2003) fit is
%still sufficiently accurate (see Fig.~\ref{fig:opticalLF}) for our
%purpose, also on account of the fact that the high-$z$
%contribution to the accreted mass density is anyway small.
%Integrating
Inserting such model in Eq.~(\ref{rhoac}), and integrating it up
to $z=6$, we get, for $k_{\rm bol}^{B}=11.8$, appropriate for
$L_B=(L_{\nu}\nu)_B$ with $\nu_B=6.8\times 10^{14}\,$Hz (Elvis et
al. 1994)
%$(L_{\nu}\nu)_{\odot,B}=3.31\times 10^{35} \times 10^{-0.4M_{\odot
%B}} =2.13 \cdot 10^{33} erg/s$),
and $\epsilon=0.1$:
\begin{equation}
\rho_{\rm acc}^{\rm opt}=1.4 \times 10^5 \frac {k_{\rm bol}}{11.8}
M_{\odot}/\hbox{Mpc}^3 \ . \label{rhoopt}
\end{equation}
with objects at $z \le 2.2$ contributing $\rho_{\rm acc}^{\rm
opt}=0.8 \times 10^5 M_{\odot}/\hbox{Mpc}^3$. Using the Boyle et
al. (2000) LF, which is however inconsistent with high redshift
data, $\rho_{\rm acc}^{\rm opt}$ increases by 20$\%$.
%{\bf It is worth noticing that integrating the Croom et al. (2003)
%LF up to $z=2.2$ we obtain a total mass density of $\rho_{\rm
%acc}^{\rm opt}=0.8 \times 10^5 M_{\odot}/\hbox{Mpc}^3$ while
%integrating the double power-law fit of Fan et al. (2001) from
%$z=2.2$ to $z=6$ and $M_B<-25$ we get $\rho_{\rm acc}^{\rm
%opt}=0.6 \times 10^5 M_{\odot}/\hbox{Mpc}^3$. The mass
%contribution from high-z, faint QSO is small. As recently reviewed
%by Osmer (2003) the space density of QSO rapidly declines for
%redshift $z>3$ and a corresponding faint-end flattening of the
%high-z optical LF is observed (Pei 1995; Cristiani et al. 2004).}
Thus the mass density accreted on BHs powering the optical QSO
emission is a factor $\simeq 3$ lower than the estimated local
SMBH mass density.

The estimate of $\rho_{\rm acc}^{\rm opt}$ is affected by
uncertainties on $k_{\rm bol}^B$ and on $\epsilon$. An upper limit
of $k_{\rm bol}^{B}=16$ can be derived from the Elvis et al.
(1994) sample. On the other hand, recent data point to a lower
bolometric correction than used in Eq.~(\ref{rhoopt}). For
instance, McLure \& Dunlop (2003) find $k_{\rm bol}^B \sim 8$, and
Vestergaard (2003) finds $k_{\rm bol}^B=9.7$ for higher redshift
QSOs. On the whole we attribute to $k_{\rm bol}^B$ an uncertainty
of about 30$\%$. It is worth noticing that no dependence of the
bolometric correction on optical luminosity has been reported.

The efficiency $\epsilon$ of conversion of accreted mass into
outgoing photons can be as high as $\simeq 0.4$ for extreme-Kerr
BHs. On the other hand, no firm lower limit to $\epsilon$ can be
set; in extreme cases a BH can grow without radiating any photon
at all. However, the low value of $\rho_{\rm acc}^{\rm opt}$ does
not necessarily imply a low value of $\epsilon$, since an
additional important contribution to the local BH mass density is
expected from highly absorbed hard X-ray selected AGNs,
contributing a large fraction of the X-ray background energy
density, but only marginally represented in optical surveys.

\begin{figure}
\begin{center}
%FDM_Acc_z_Croom_Ueda.ps
\includegraphics{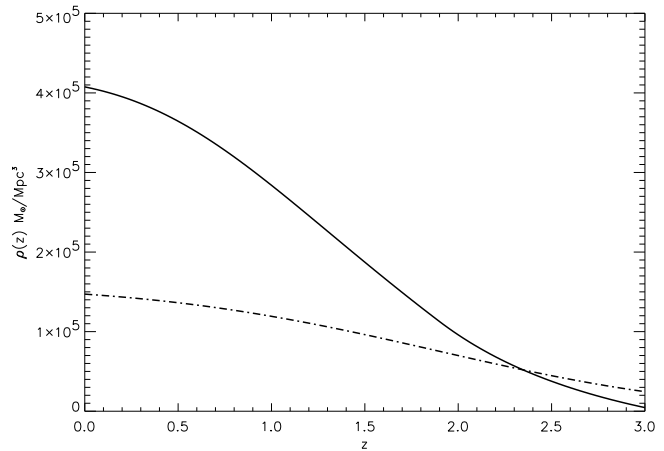} \vspace{6.4cm} \caption{Accreted mass density as a
function of redshift. The solid and the dot-dashed lines show the
increase, with decreasing redshift, of the comoving accreted mass
density as inferred from the epoch-dependent hard X-ray luminosity
function by Ueda et al. (2003), with a luminosity dependent
bolometric correction (see text), and from the optical luminosity
function by Croom et al. (2003), respectively.} \label{rhohistory}
\end{center}
\end{figure}

\subsection{Mass accreted on X-ray selected AGNs}

Recent data (see Hasinger 2003 for a review) have shown that a
large fraction of the hard X-ray background is contributed by
partially or completely covered AGNs in the luminosity range
$\log(L_{2\hbox{-}10{\rm keV}})=42-44$ erg/s and at redshifts
$z=0.5$--1. A comprehensive study of the redshift-dependent hard
X-ray AGN LF, including the  distribution of the absorption column
density N$_{H}$, has been recently carried out by Ueda et al.
(2003; U03 from now on). In the following we will refer to their
LDDE model, with the additional  fraction of AGN with $24 <
\log(N_H) < 25$ required in order to fit the XRB with the most
recent normalizations (Vecchi et al. 1999; Barcons et al 2000;
Gilli 2003). The additional AGN fraction implies an increase of
the mass density by 25$\%$.

% in the following we will refer to their LDDE
%model.

%Inserting the above mentioned bolometric correction $k_{\rm bol}(2-10
%keV)\simeq 30$, eq. 9 specialized to the hard X-ray AGN yields
%\begin{equation}
%\rho_{\rm BH}^{XAGN}\simeq 2.8 (3?) \times 10^5 M_{\odot}/Mpc^3.
%\end{equation}

%Check please! The value is about 30$\%$ higher that that found
%above from the intensity of the HXRB; in fact the U03 model
%predicts an intensity at the level measured by Vecchi et al (??).
%On the other hand the agreement between the two estimates is
%encouraging.

Unfortunately the available information on the overall spectral
energy distribution of hard X-ray selected objects (and
particularly of the faint ones, which are the most relevant to
estimate the low mass end of the MF) is scanty, so that estimates
of the bolometric corrections are difficult. The bolometric
correction, $k_{\rm bol}^{2-10}\simeq 32$, derived by Elvis et al.
(1994), refers to optically bright quasars. Evidences for an
increase of the hard X-ray to optical luminosity ratio,
$L_{HX}/L_{\rm opt}$ with decreasing optical luminosity have been
reported by Vignali et al. (2003) and bolometric corrections,
$k_{\rm bol}^{2-10}\simeq 12-18$, substantially smaller than the
Elvis et al. (1994) value, have been estimated at least for a few
Seyfert galaxies (Fabian 2003). Moreover, in order to match the
optical LF of Boyle et al. (2000) starting from the hard X-ray LF,
U03 had to assume  that $L_{2\ keV} \propto L_{2500A} ^{0.7}$ in
close agreement with the observational data by Vignali et al.
(2003).

The luminosity dependence of the X-ray to optical luminosity ratio
does not necessarily imply a higher efficiency of low luminosity
objects in producing X-ray (compared to optical) photons. An
alternative explanation, borne out by evidences of larger covering
factors for Seyfert galaxies compared to QSOs, is stronger dust
extinction for lower luminosity objects which are less capable of
pushing away the surrounding medium, as suggested long ago (Cheng
et al. 1983).

If the optical/UV bolometric correction is independent of
luminosity, the U03 relationship between UV and X-ray luminosity
implies:
\begin{equation}
k_{\rm bol}^{2-10}=17
\left(\frac{L_{2-10}}{10^{43}\hbox{erg}\,\hbox{s}^{-1}}
\right)^{0.43}. \label{kbolX}
\end{equation}
Inserting Eq.~(\ref{kbolX}) in Eq.~(\ref{rhoac}), assuming
$\epsilon/(1-\epsilon)=0.1$, and integrating over the luminosity
and redshift intervals ($41.5 \leq \log(L_{2-10 {\rm keV}}) \leq
46.5$ and $z\leq 3$) investigated by Ueda et al. (2003), we find
\begin{equation}
\rho_{\rm acc}^{HX}\simeq 4.1 \times 10^5  M_{\odot}/\hbox{Mpc}^3
\ . \label{rhoX}
\end{equation}
If we extrapolate the LF  up to  $z=6$, we get a mass density
larger by 15$\%$.

As a consistency check, we have subtracted the contribution of
Type 2 AGNs, following the prescriptions given by U03, in order to
get the contribution to the local mass density of Type 1 objects
only. We find:
\begin{equation}
\rho_{\rm acc}^{{\rm Type}\,1}\simeq 1.5 \times 10^5
M_{\odot}/\hbox{Mpc}^3 \ , \label{Type1}
\end{equation}
in close agreement with the result obtained using the optical LF
[Eq.~(\ref{rhoopt})]. The relatively large contribution of the
optically selected AGNs to the local BH mass density ($> 30\%$),
despite their small ($<20\%$) contribution to the intensity of the
HXRB, reflects their lower X-ray to optical luminosity ratio.
Since $\rho_{\rm acc}^{{\rm Type}\,1} \propto
[(1-\epsilon)/\epsilon] k_{\rm bol}^{2-10}$ and $\rho_{\rm
acc}^{\rm opt}\propto [(1-\epsilon)/\epsilon] k_{\rm bol}^{B}$,
from the agreement between the two estimates we can conclude that
the uncertainty on $k_{\rm bol}^{2-10}$ is similar to that on
$k_{\rm bol}^B$,  i.e. $\simeq 30\%$.

The mass accretion history is illustrated by Fig.~\ref{rhohistory}
showing the increase with decreasing redshift of the comoving
accreted mass density, $\rho_{\rm acc}(z)$, as inferred from the
optical (dot-dashed line) and from the hard X-ray (solid line)
epoch dependent comoving luminosity function $n(L,z)$:
\begin{equation}
\rho_{\rm acc}(z)=\frac {1-\epsilon}{\epsilon c^2}
\int_{z}^{z_{max}}\!\!\!\!\!\!\!\! dz'
\frac{dt}{dz'}\int_{L_{min}}^{L_{max}}\!\!\!\!\!\!\!\! dL'\,
k_{\rm bol} L'\, n(L',z)  ,
\end{equation}
where the X-ray (but not the optical) bolometric correction is a
function of luminosity, as discussed above. As shown by
Fig.~\ref{rhohistory}, most of the accretion occurs at $z>1.5$ for
optically selected AGNs, and at $z<1.5$ for hard X-ray selected
AGNs.

%Not integrating in redshift in equation 9, we can get information
%on how the accreted matter piles up in BHs in time. It is apparent
%that most of the accreted matter in optically selected AGN are
%already at palce at $z>1.5$, while most of the accretion on hard
%X-ray selected AGN occurs at $z<1.5$.

The close correspondence of the accreted mass density inferred
from the hard X-ray LF with the local SMBH mass density $\rho_{\rm
BH}\simeq 4.2\pm 1.0 \times 10^{5}\, M_{\odot}/\hbox{Mpc}^3$ (see
Table~\ref{rhoBH}) for $\epsilon/(1-\epsilon)=0.1$ shows that {\it
there is no much room for really ``dark'' accretion} (i.e. for
accretion with radiative efficiency $\epsilon \ll 0.1$),
confirming the findings by Salucci et al. (1999) and of Marconi et
al. (2004), unless the luminous phases of the AGNs are
characterized by radiative efficiencies much higher than the
usually adopted value. But even if $\epsilon$ is close to the
maximum allowed values ($\simeq 0.3$--0.4; Thorne 1974) the
accreted mass accounts for $\geq 25$--$30 \%$ of the local SMBH
mass density, and one would be left with the problem of accounting
for the correlations between $M_{\rm BH}$ and the bulge mass or
velocity dispersion which arise naturally as a consequence of
feedback associated to radiative accretion (Silk \& Rees 1998;
Cavaliere et al. 2002; King 2003; Granato et al. 2004).

A more explicit test of the role of accretion is obviously the
comparison, presented in the next Section, of the resulting MF
with the local SMBH MF.

\begin{figure}
\begin{center}
\includegraphics{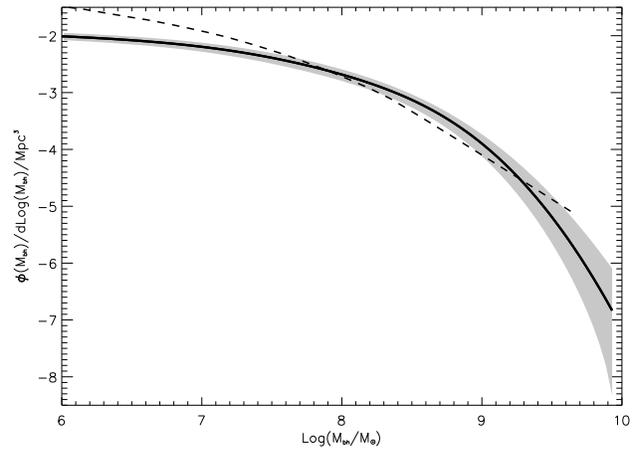} \vspace{6.4cm} \caption{Local SMBH MF
(solid line), including SMBHs hosted by both early- and late-type
galaxies, with its $1\sigma$ uncertainty (shaded area), compared
with the accreted MF (dashed line) estimated from the X-ray LF by
U03, using a luminosity dependent bolometric correction. Such
estimate is obtained by differentiating the integral mass function
[Eq.~(\protect\ref{YuLu1})] with $\lambda=L/L_{\rm Edd}=1$.}
\label{estMF1}
\end{center}
\end{figure}

\begin{figure}
\begin{center}
%\special{psfile=./MF_tot.ps hscale=50 vscale=50 hoffset=-40
\includegraphics{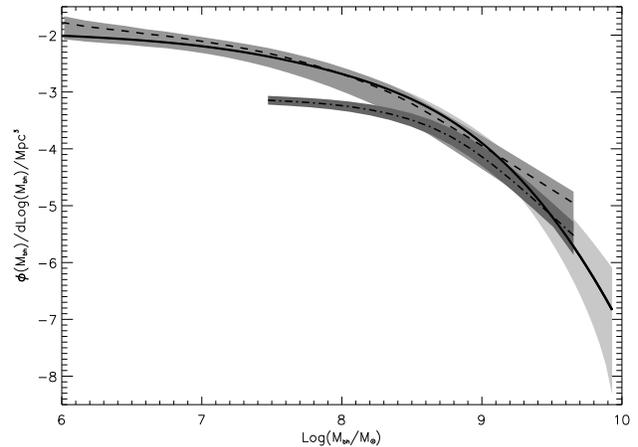} \vspace{6.4cm} \caption{{Comparison of the accreted
MF (dashed line) computed as in Fig.~\protect\ref{estMF1}, but for
$\lambda$ given by Eq.~(\protect\ref{lambda}) with the local SMBH
MF (solid line, with $1\sigma$ uncertainties represented by the
shaded area). The dot-dashed line shows the accreted MF of
optically selected QSOs ($M_B< -22.5$).}} \label{estMF2}
\end{center}
\end{figure}

\begin{figure}
\begin{center}
%\special{psfile=./CMF_agn_12.ps hscale=50 vscale=50 hoffset=-40
\includegraphics{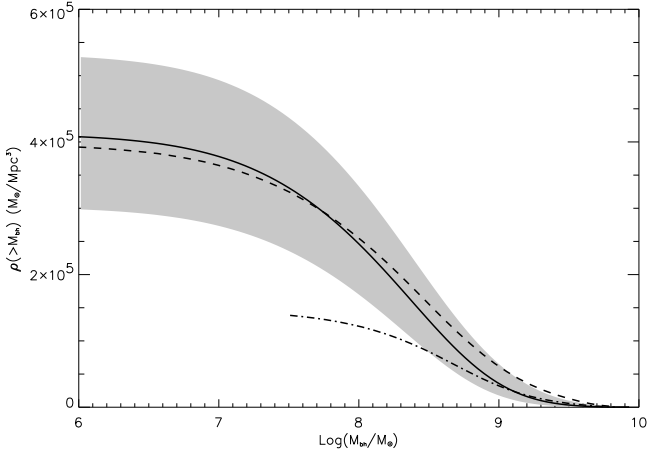} \vspace{6.4cm} \caption{Cumulative local SMBH MF
(dashed line) with its 1$\sigma$ uncertainties, compared with the
cumulative MF of  optically selected QSOs ($M_B< -22.5$;
dot-dashed line) plus X-ray selected Type 2 AGNs.} \label{cumMF}
\end{center}
\end{figure}

\begin{figure}
\begin{center}
\includegraphics{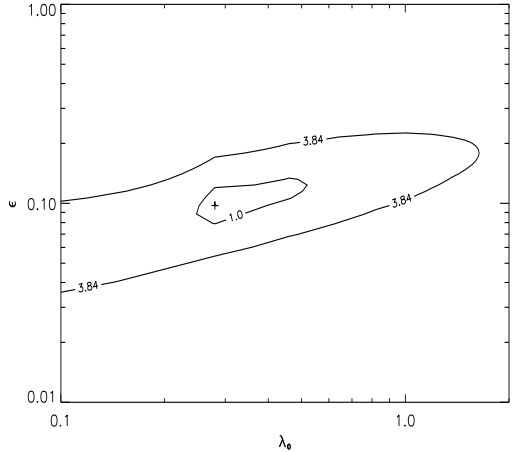} \vspace{6.4cm} \caption{Iso-$\chi^2$ contours in the
$\epsilon$--$\lambda_0$ plane for the match between the accreted
and the local SMBH MFs. The contours are labelled with their value
of $\Delta \chi^2$. The projections of the $\Delta \chi^2=1$ and
$\Delta \chi^2=3.84$ contours on the axis corresponding to a
parameter give the 68\% and 95\% confidence intervals,
respectively, for such parameter.} \label{contour}
\end{center}
\end{figure}

\section{The local accreted mass function}

The AMF can be derived from the AGN LF once a relationship between
luminosity and BH mass is established (Chokshi \& Turner 1992).
Salucci et al. (1999) compared, under plausible assumptions, the
accreted MF with the local SMBH MF to infer information on the
accretion history. This point has been recently reexamined by a
number of authors (e.g. Yu \& Tremaine 2002; Aller \& Richstone
2002; McLure \& Dunlop 2003; Yu \& Lu 2004; Marconi et al. 2004),
who reached different conclusions.

Let us assume that the local SMBH mass is mostly due to radiative
accretion and that the accretion rate $\dot{M}_{\rm BH}$ is
proportional to $M_{\rm BH}$ (see e.g. Small \& Blandford 1992;
Cavaliere \& Vittorini 2002; Marconi et al. 2004), at least during
the main accretion phases. A recent analysis of SDSS quasars
suggests that the two quantities are correlated (McLure \& Dunlop
2003), although with a huge scatter, at least partly due to the
uncertainties in BH mass estimates. An almost constant
$\dot{M}_{\rm BH}/M_{\rm BH}$ is also expected, according to the
physical model of Granato et al (2004), during the fast growth of
the SMBHs and up to the bright quasar phase.

If $\dot{M}_{\rm BH}/M_{\rm BH}$ is constant, the bolometric
luminosity grows exponentially (as does the BH mass):
\begin{eqnarray}
L_{\rm bol}(t)&=&\epsilon \dot{M}_{\rm acc}
c^{2}=\frac{\epsilon}{1-\epsilon} \dot{M}_{\rm BH} c^{2} \nonumber
\\ &=&\frac{\lambda c^2}{t_{E}}M_{\rm BH}(t_{i})\exp
\left[\frac{(t-t_{i})}{t_{ef}}\right], \label{Lt}
\end{eqnarray}
with e-folding time
\begin{equation}
t_{ef}=\frac{\epsilon t_{E}}{(1-\epsilon)\lambda}, \label{tef}
\end{equation}
where $\lambda$ is the average ratio $L/L_{\rm Edd}$, $t_E$ is the
Eddington time and $t_i$ is the time when the growth starts. The
e-folding time equals the Salpeter time if $\lambda=1$.

The growth stops when the SMBH reaches its maximum mass, set equal
to its present-day mass $M_{\rm BH}^{0}$, i.e. we neglect the mass
increase during the declining phase of the light curve (see Yu \&
Lu 2004). The maximum bolometric luminosity is then:
\begin{equation}
L_{\rm bol,max}({M}_{\rm BH}^0)=\lambda \frac{M_{\rm BH}^{0}}{t_E}
c^2. \label{Lmax}
\end{equation}
Under these assumptions, the local SMBH MF is related to the
epoch-dependent LF in a given observational band by the energy
balance equation:
\begin{eqnarray}
& &\! \!\!\!\!\!\! \!\!\!\!\!\! \frac {1-\epsilon}{\epsilon c^2}
\int_{0}^{t_0}dt\int_{\bar{L}}^{\infty}dL k_{\rm bol} L n(L,t)=
\nonumber \\
& = & \int_{\bar{M}_{\rm BH}^0}^{\infty}dM_{\rm BH}^{0} n(M_{\rm
BH}^{0}) \left[M_{\rm BH}^0-\bar{M}_{\rm BH}^0 \right] ,
\label{YuLu1}
\end{eqnarray}
where ${\bar{L}}=L_{\rm max}({\bar{M}}_{\rm BH}^0)$. The local MF
$n(M_{\rm BH}^{0})$ is straightforwardly obtained differentiating
Eq.~(\ref{YuLu1}) with respect to ${\bar{M}_{\rm BH}^0}$. It is
worth noticing that the above equation can be also derived
starting from the continuity equation (see e.g. Yu \& Lu 2004).

In Fig.~\ref{estMF1} the estimated AMF derived from the
epoch-dependent X-ray LF by U03, assuming Eddington limited
accretion ($\lambda=1$), is compared with the local SMBH MF
(including SMBHs hosted by both early- and late-type galaxies).
Although the two curves are rather close to each other, their
shapes differ. The fact that the assumption of Eddington limited
accretion lead to an AMF exceeding the SMBH MF in some mass range
shows that it cannot be true for all epochs and/or luminosities.
Indeed, low-$z$/low luminosity AGNs are known to be radiating well
below the Eddington limit (Wandel et al. 1999), and recent
estimates suggest that quasars are in a sub-Eddington regime up to
$z \simeq 2$ (McLure \& Dunlop 2003; Vestergaard 2003).

The match between the AMF and the local SMBH MF indeed improves
significantly (Fig.~\ref{estMF2}) if we adopt a redshift dependent
Eddington ratio of the form:
\begin{equation}
\lambda(z)= \left\{ \begin{array}{ll} \lambda_0 & \mbox{if $z \ge
3$} \nonumber \\ & \\ \lambda_0 [(1+z)/4]^{\alpha} & \mbox{if $z <
3$}
\end{array} \right. , \label{lambda}
\end{equation}
with $\lambda_0 =1$ and $\alpha=1.4$. The discrepancy at $M_{\rm
BH}\geq 10^{9}$ $M_{\odot}$ is only marginally significant being
at slightly more than $1\sigma$ level. On the other hand, the
generally higher AMF estimate derived from the X-ray, compared to
that from the optical, LF (see Fig.~\ref{estMF2}), reflects the
strong luminosity dependence of the fraction of Type 2 AGNs, which
are represented in the X-ray, but not in the optical, LF. X-ray
surveys (see, e.g., Hasinger 2003) have shown that the Type 2
fraction increases from $\simeq 30\%$ at high luminosities
($L_{2-10\,{\rm keV}}\gsim 10^{44}\,\hbox{erg}\,\hbox{s}^{-1}$) to
$\simeq 70$--$80\%$ at low luminosities ($L_{2-10\,{\rm keV}}\lsim
3\times 10^{42}\,\hbox{erg}\,\hbox{s}^{-1}$), consistent with the
results of optical spectroscopic surveys of complete samples of
nearby galaxies, without pre-selection (Huchra \& Burg 1992; Ho et
al. 1997). As a check, we have computed and plotted in
Fig.~\ref{cumMF} the cumulative accreted mass density function
obtained summing the contribution of the optically selected QSOs
to that of Type 2 X-ray selected AGNs; again, the agreement with
the local SMBH mass density function is very good.

We checked that a dependence of $\lambda$ on luminosity, as
suggested by Salucci et al. (1999), rather than on redshift,
yields an equally good fit. %The number density of SMBHs with
%$M_{\rm BH}>10^{6}$ is $n_{SMBH}\simeq 1.7 \times 10^{-2}$
%$Mpc^{-3}$, quite close to that derived from the LMF. The BHs that
%powered X-ray selected (possibly type 2) AGN outnumber the BHs
%associated to optically selected QSOs by a factor larger than 5 at
%$M_{\rm BH}\leq 3\times 10^7$ $M_{\odot}$, again in agreement with
%the findings of Salucci et al (1999).

Requiring that the AMF matches the local SMBH MF we obtain
constraints on the radiative efficiency and on the maximum value
of the Eddington ratio [Eq.~(\ref{lambda})]. A minimum $\chi^2$
analysis yields $\epsilon \simeq 0.09\,(+0.04,-0.03)$ and
$\lambda_0\simeq 0.3\,(+0.3,-0.1)$ (68\% confidence errors; see
Fig.~\ref{contour}). The constraints on the parameter $\alpha$
ruling the evolution of the Eddington ratio [Eq.~(\ref{lambda})]
are rather loose ($0.3 \le \alpha \le 3.5$).

\begin{figure}
\begin{center}
\includegraphics{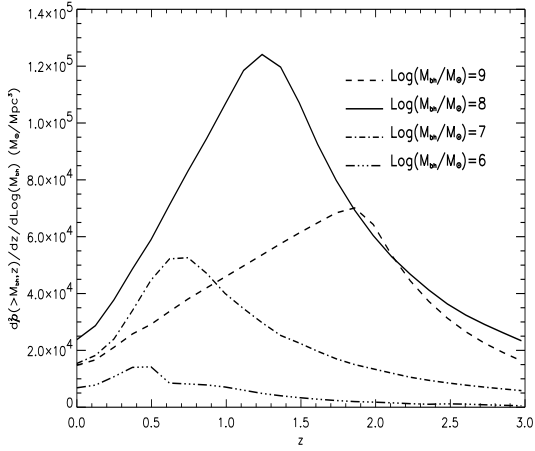} \vspace{6.4cm} \caption{Contributions to the local
SMBH MF as a function of redshift, for several values of $M_{\rm
BH}^{0}$.} \label{MAR}
\end{center}
\end{figure}

\section{Accretion history and AGN visibility times}

Replacing $t_0$ with $t$ in Eq.~(\ref{YuLu1}) and differentiating
with respect to $z$ and $M_{\rm BH}^0$ we get the contributions to
the $n(M_{\rm BH}^0)$ from different cosmic epochs, shown for
several values of $M_{\rm BH}^0$ in Fig.~\ref{MAR}, which
evidences that mass is accreted earlier and more rapidly by the
more massive BHs.

%Information on timescale of the luminous activity  can be
%recovered by the condition
%that the AGN visibility timescales determined via  the LF and via the
%local SMBH mass function are equal:
%
%\begin{eqnarray}
%& &\! \!\!\!\!\!\! \!\!\!\!\!\!
%\int_{0}^{t_0}dt\int_{\bar{L}}^{\infty}dL  n(L,t) \nonumber  \\
%&=& \int_{\bar{M}_{\rm BH}^{0}}^{\infty}dM_{\rm BH}^{0}
%n(M_{\rm BH}^{0}) \tau_{lum}(M_{\rm BH}^{0}),
%\label{tauvis1}
%\end{eqnarray}
%
%where $\tau_{lum}(M_{\rm BH}^{0})$ is the time spent at $L\geq
%\bar{L}$. This equation follows from continuity equation, as noted by Yu
%\& Tremaine (2002).

%$$
%\int_{0}^{t_0}dt\int_{\bar{L}}^{\infty}dL  n(L,t)=
%\int_{\bar{M}_{bh}^{0}}^{\infty}dM_{bh}^{0} \times
%$$
%$$
%n(M_{bh}^{0}) \int_{\bar{L}}^{\infty}dL  \tau (L \geq \bar{L} \mid
%M_{bh}^{0})(M_{bh}^{0})P(L\mid M_{bh}^{0}). \eqno{(16b)}
%$$

The the mean time spent by an AGN at  $L\geq \bar{L}$, averaged
over the MF, writes:
\begin{equation}
\langle \tau_{\rm lum}(\bar{L}) \rangle =
\frac{\int_{0}^{t_0}dt\int_{\bar{L}}^{\infty}dL
n(L,t)}{\int_{\bar{M}_{\rm BH}^{0}}^{\infty}dM_{\rm BH}^{0}
n(M_{\rm BH}^{0})} . \label{taulum}
\end{equation}
As noted by Yu \& Tremaine (2002), $\langle\tau_{lum}\rangle$ is
independent of $\epsilon$. It depends however on $k_{\rm bol}$ and
on $\lambda$ since
\begin{equation}
\bar{L}=\frac{\lambda}{k_{\rm bol}} \frac{\bar{M}_{\rm
BH}^{0}}{t_E} c^2 . \label{Lbar}
\end{equation}
Adopting the U03 LF, we find $\langle \tau_{\rm lum}\rangle \simeq
1.5\times 10^8\,$ yr if we adopt a redshift-dependent Eddington
ratio [Eq.~(\ref{lambda})] and $\simeq 8\times 10^7\,$ yr if we
keep $\lambda=\lambda_0$.

This is however a lower limit to the time interval $\tau_{\rm
vis}$, spent at $L\geq L_{\rm min}$, the minimum luminosity
included in the LF. If $\dot{M}/M_{\rm BH}=\hbox{const}$, the
duration of the visible phase is simply:
\begin{equation}
\tau_{\rm vis}(M_{\rm BH}^0)=t_{ef}\ln \left[\frac{L_{\rm
max}(M_{\rm BH}^0)}{L_{\rm min}}\right]= t_{ef}\ln
\left[\frac{M_{\rm BH}^0}{M_{\rm BH}^{\rm min}}\right],
\label{tauAGN}
\end{equation}
where $M_{\rm BH}^{\rm min}$ is the BH mass when radiative
accretion yields luminosity $L \geq L_{\rm min}$. In the case of
the X-ray LF of U03, the minimum 2--10 keV luminosity included at
$z\simeq 0.8$, where the contribution to the X-ray background
peaks, is $\log(L_{\rm min})\simeq 42$--42.4. With the bolometric
correction given by Eq.~(\ref{kbolX}), the corresponding minimum
BH mass is:
\begin{equation} M_{\rm BH}^{\rm min} = \frac {5\times
10^4}{\lambda} \frac {L_{\rm min}}
{10^{42}\hbox{erg}\,\hbox{s}^{-1}} \ M_{\odot}. \label{MbarX}
\end{equation}
The minimum BH mass contributing to the LF of optically selected
QSOs for the bolometric correction by Elvis et al. (1994) is :
\begin{equation}
M_{\rm BH}^{\rm min}= \frac {3\times 10^7}{\lambda}
10^{-0.4(22.5+M_B^{\rm max})} \ M_{\odot}, \label{Mbaropt}
\end{equation}
where $M_B^{\rm max} \simeq -22.5$ (Boyle et al. 2000; Croom et
al. 2003).

The condition that the AGN visibility timescales determined via
the LF and via the local SMBH mass function are equal:
\begin{eqnarray}
& &\! \!\!\!\!\!\! \!\!\!\!\!\!
\int_{0}^{t_0}dt\int_{\bar{L}}^{\infty}dL  n(L,t) \nonumber  \\
&=& t_{ef}\int_{\bar{M}_{\rm BH}^{0}}^{\infty}dM_{\rm BH}^{0}
n(M_{\rm BH}^{0}) \ln(\frac{M_{\rm BH}^0}{\bar{M}_{\rm BH}^0}),
\label{tauvis}
\end{eqnarray}
add further constraints on $\epsilon$ and $\lambda_0$. Using the
U03 LF and assuming a 30\% uncertainty on the bolometric
correction for X-ray luminosities, we find an allowed range for
$\epsilon$ pretty similar to that following from the match between
the AMF and the local SMBH mass function ($0.06\leq \epsilon \leq
0.13$) while the constraints on the Eddington ratio are looser
($0.5 \leq \lambda_0 \leq 2$).

\section{Discussion and conclusions}

The analysis reported in the first part of the paper shows that
the local mass function of SMBHs can be rather accurately
assessed. More in detail, the MF of SMBHs hosted in early-type
galaxies can be obtained exploiting the velocity dispersion or
luminosity functions of host galaxies, coupled with the $M_{\rm
BH}$--$\sigma$ or $M_{\rm BH}$--$L_{\rm sph}$ relationships,
respectively. The results obtained in the two ways are in
remarkable agreement with small uncertainties up to $M_{\rm
BH}\geq 10^9\, M_{\odot}$ (cfr. Fig.~\ref{SMBHMF2}). The
contribution from SMBHs hosted by late-type galaxies is more
uncertain, and is mostly confined to the low mass end of the MF.

The overall SMBH mass density amounts to $\rho _{\rm
BH}^{0}=(4.2\pm 1.1) \times 10^5 M_{\odot}/\hbox{Mpc}^3$, with a
contribution from SMBHs in late-type galaxies of $\simeq 25\%$.
This value of $\rho _{\rm BH}^{0}$ is higher than those found by
Yu \& Tremaine (2002), by Aller \& Richstone (2002) and by McLure
\& Dunlop (2003) (who have not considered the contribution from
SMBH residing in late type galaxies), but is in excellent
agreement with the results by Marconi et al. (2004). The local
number density of the SMBHs more massive than $10^6\, M_\odot$ is
$n(M_{\rm BH}^0> 10^6 \ M_{\odot})\simeq 1.7 \times
10^{-2}\,\hbox{Mpc}^{-3}$, which corresponds to the number of
bulges and spheroids with $M_{\rm sph}>5 \times 10^8$ $M_{\odot}$.

The Soltan (1982) argument applied to the hard X-ray selected
AGNs, allowing for a luminosity-dependent bolometric correction
(U03; Fabian 2003), yields,  for a mass to radiation conversion
efficiency $\epsilon=0.1$, an accreted mass density of $\rho_{\rm
acc}^{HX}\simeq 4.1 \times 10^5 M_{\odot}/\hbox{Mpc}^3$, in close
agreement with the local SMBH mass density, indicating that most
of the BH masses were accumulated by radiative accretion, as
previously concluded by Salucci et al. (1999). Optically selected
QSOs account for only $\simeq 35\%$ of the total SMBH mass
density. The dominant contribution comes from Type 2 AGNs, mostly
missed by optical surveys.

Not only the mass density, but also the MFs of the SMBHs and of
the accreted mass match remarkably well, if we allow for a
decrease of the Eddington ratio $\lambda=L/L_{\rm Edd}$ with
redshift [Eq.~(\ref{lambda})], as suggested by observations
(McLure \& Dunlop 2003; Vestergaard 2003). Optically selected,
Type 1 AGNs account for the high mass tail of the AMF, while Type
2 AGNs take over at lower masses, reflecting the strong increase
with decreasing luminosity of the Type 2 to Type 1 ratio
demonstrated by hard X-ray surveys (see, e.g., Hasinger 2003) and
consistent with the outcome of spectroscopic surveys of complete
samples of nearby galaxies, without pre-selection (Huchra \& Burg
1992; Ho et al. 1997).

The alternative possibility that most of the mass has been
accumulated by ``dark'' accretion (i.e. accretion undetectable by
either optical or hard X-ray surveys, as in the case of BH
coalescence), is severely constrained by the above results. In
order to make room for this possibility one has to assume that the
radiative efficiency during the visible AGN phases is at the
theoretical maximum of $\epsilon\simeq 0.3$--0.4. But even in this
case (unless the bolometric correction is far lower than currently
estimated) the contribution of radiative accretion to the local
SMBH mass density is $\gsim 25\%$, and one is left with the
problem of fine tuning the radiative and non-radiative
contributions in order not to break down the match with the local
SMBH MF obtained with radiative accretion alone. One would also
face the problem of accounting for the tight relationships between
BH mass and mass or velocity dispersion of the spheroidal host,
naturally explained by feedback associated to radiative accretion.
For these reasons it is unlikely that the present day SMBH mass
function has been built {\itshape mostly} through 'dark' accretion
or merging of BHs.

%Very similar results have been recently claimed by Bromley et
%al. (2004) who studied simulations of SMBH growth in dark matter
%halos. They find that to match the number of SDSS quasars at
%redshift 6 and in order not to overproduce the local SMBH mass
%density, a severe lowering of the efficiency in the SMBH growth
%through merging processes must be accounted for. Specifically they
%find that merging could be effective for building up SMBH masses only
%up to a value close to $10^6 M_{\odot}$.

If indeed the SMBH MF has to be accounted for by radiative
accretion, the requirement that it fits together with the AMF
constrains the radiative efficiency and the maximum Eddington
ratio to $\epsilon \simeq 0.09\,(+0.04,-0.03)$ and
$\lambda_0\simeq 0.3\,(+0.3,-0.1)$ (68\% confidence errors). The
condition that the mean AGN visibility timescale computed via the
LF and via the local MF are equal [Eq.~(\ref{tauvis})] yields an
allowed range for $\epsilon$ very close to the above ($0.06\leq
\epsilon \leq 0.13$) and looser (but consistent) constraints on
$\lambda_0$.

The analysis of the accretion history highlights that the most
massive BHs (associated to bright optical QSOs) accreted their
mass faster and at higher redshifts (typically at $z>1.5$), while
the lower mass BHs responsible for most of the hard X-ray
background have mostly grown at $z<1.5$ (see
Figs.~\ref{rhohistory}  and \ref{MAR}).  The different
evolutionary behaviour of the two AGN populations can be
understood if gas accretion is regulated by star formation and by
feedback both from supernova explosions and nuclear activity (e.g.
Kawakatu \& Umemura 2003; Granato et al. 2004). In this framework
it is expected that the hosts of the most massive BHs have the
oldest stellar populations (Cattaneo \& Bernardi 2003).

The mass-weighted duration of the luminous AGN phase is found to
be $\langle\tau_{\rm lum}\rangle\simeq 0.5$--$1.5 \times
10^8\,$yr. Yu \& Tremaine (2002), using a similar method, got
slightly lower values, because they used in Eq.~(\ref{taulum}) the
optical QSO LF and the local MF of SMBHs in early-type galaxies,
which, as shown in Sect.~3, is only partly accounted for by
optically selected QSOs. Yu \& Lu (2004), from a detailed
modelling the luminosity evolution of individual QSOs, derived a
lower limit $\tau_{\rm lum}\geq 4\times 10^7\,$yr.

If the accretion rate per unit BH mass, $\dot{M}_{\rm BH}/M_{\rm
BH}$, is constant (as in the case of Eddington limited accretion)
during the main accretion phases, the visibility times increase
with BH mass [Eq.~(\ref{tauAGN})], consistent with the finding  by
McLure \& Dunlop (2003) that the ratio of SMBH in their optically
selected sample at $z\simeq 2$ to the corresponding number density
at the present day increases with BH mass. These authors estimate,
for the most massive BHs ($M_{\rm BH} \ge 10^{9.5}\, M_\odot$), a
lower limit to $\tau_{\rm vis}$ (lifetime in their terminology) of
$10^8\,$yr.

%Adopting the redshift-dependent Eddington ratio Setting
%[Eq.~(\ref{lambda})]
Setting $\lambda=1$ and $\epsilon/(1-\epsilon)=0.1$ and inserting
in Eq.~(\ref{tauAGN}) the value of $\bar{M}_{\rm BH}$ given by
Eq.~(\ref{Mbaropt}) with $M_B = -22.5$, we get $\tau_{\rm
vis}(10^{9.3}\, M_\odot) \simeq 2 \times 10^8\,$yr. Inserting
instead in Eq.~(\ref{tauAGN}) the value of $\bar{M}_{\rm BH}$
given by Eq.~(\ref{MbarX}) with $\log(L) = 42$ we get $\tau_{\rm
vis}(10^{6}\, M_\odot) \simeq 1.2\times 10^8\,$yr, $\tau_{\rm
vis}(10^{8}\, M_\odot) \simeq 3\times 10^8\,$yr and $\tau_{\rm
vis}(10^{9.3}\, M_\odot) \simeq 4.4\times 10^8\,$yr. Such
visibility times increase with decreasing redshift for $z<3$
[Eq.~(\ref{lambda})].

On the contrary, Marconi et al. (2004) infer mean ``lifetimes''
increasing with decreasing present day BH masses. For $\lambda=1$
and $\epsilon=0.1$ they get $\sim 1.5\times 10^8\,$yr for $M_{\rm
BH}^0 > 10^9\, M_\odot$ and $\sim 4.5\times 10^8\,$yr for $M_{\rm
BH}^0 < 10^8\, M_\odot$. However, the latter ``lifetime''
corresponds to $\simeq 11$ e-folding times, i.e. to a mass
increase by a factor $4.5 \times 10^4$. But then for a large
fraction of their ``lifetime'' low mass BHs are too faint to be
included in actual LFs [cf. Eq.~(\ref{MbarX}) and
Eq.~(\ref{Mbaropt})].

In this context we also note that the amount of time spent above
the threshold for inclusion in the LF by objects reaching large
luminosities/masses cannot be constrained by the LFs themselves
since such objects are too rare to contribute significantly to the
faint end of the LF.

Our estimates for $\tau_{\rm vis}$ are within the broad
constraints imposed on one side by the QSO evolution timescale ($<
1\,$Gyr) and on the other side by the proximity effect ($>
10^4\,$yr; see Martini 2004). The low ``lifetimes'' inferred
(although with large uncertainties) from clustering properties of
optical QSOs (Martini \& Weinberg 2001; Martini 2004) depend
basically on the crude assumption that as soon as a massive halo
virializes, the QSO appears, without any delay.

Short  visibility times ($\tau_{\rm vis}\leq 0.02\,$Gyr) are also
implied by the models of QSO LFs presented by Haehnelt, Natarajan
\& Rees (1998), and Wyithe \& Loeb (2003). However, Hosokawa
(2002) has shown that, in the same general framework, the QSO LF
can be reproduced with substantially higher values of $\tau_{\rm
vis}$. Again, short timescales mainly follow from the assumption
of immediate QSO ignition at the virialization of the host DM
halo. As shown and discussed by Monaco et al. (2000) and Granato
et al. (2001), a delay between virialization and AGN ignition is a
key ingredient to understand the QSO LF and the relationship
between evolutionary histories of QSOs and of the massive
spheroidal galaxies hosting them. Granato et al. (2004) have
presented a detailed physical model quantifying such delay and its
increase with decreasing mass.

\section*{ACKNOWLEDMENTS}
We thank P. Monaco, P. Tozzi, A. Marconi and A. Cavaliere for
helpful discussions. We are very grateful to the referee for
perceptive comments that helped significantly improving the paper.
Work supported in part by a MIUR/COFIN grant.

\end{document}


\begin{thebibliography}{}  % as in Monthly Notices
\bibitem[Aller \& Richstone 2002]{all02} Aller M. C., Richstone D., 2002, AJ, 124, 3035
\bibitem[]{} Barcons X., Mateos S., Ceballos M.T., 2000, MNRAS, 316,
L13-L16
\bibitem[\protect\citeauthoryear{Barger et al.}{2003}]{2003ApJ...584L..61B}
Barger A.J., Cowie L.L., Capak P., Alexander D.M., Bauer F.E.,
Brandt W.N., Garmire G.P., Hornschemeier A.E., 2003, ApJ, 584, L61
%\bibitem[]{} Begelman M. C., 2003, in Carnegie Observatories Astrophysics
%Series, Vol. 1: Coevolution of Black Holes and Galaxies, ed. L.C.
%Ho (Cambridge Univ. Press), in press, astroph/0303040
\bibitem[]{} Bernardi M. et al., 2003, AJ, 125, 1817
\bibitem[]{} Blanton M. R. et al., 2001, AJ, 121, 2358
\bibitem[]{} Blanton M. R. et al., 2003, ApJ, 592, 819
\bibitem[]{} Boyle B. J., Shanks T., Croom S. M., Smith R.J. Miller L.,
Loaring N., Heymans C., 2000, MNRAS 317, 1014
\bibitem[]{} Borriello A., Salucci P., Danese L., 2003,
MNRAS, 341, 1109
%\bibitem[]{} Bromley J. M., Somerville R.
%S., Fabian A. C., 2004, MNRAS, 350, 456
\bibitem[]{} Cattaneo A., Bernardi M., 2003, MNRAS, 344, 45
\bibitem[]{} Cavaliere A., Vittorini V., 2002, ApJ, 570, 114
\bibitem[\protect\citeauthoryear{Cavaliere, Lapi, \&
Menci}{2002}]{2002ApJ...581L...1C} Cavaliere A., Lapi A., Menci
N., 2002, ApJ, 581, L1
\bibitem[\protect\citeauthoryear{Cheng, Danese, \& de
Zotti}{1983}]{1983MNRAS.204P..13C} Cheng F.Z., Danese L., de Zotti
G., 1983, MNRAS, 204, 13P
\bibitem[]{} Cole S. et al. 2001, MNRAS, 326, 255
\bibitem[]{} Chokshi A., Turner E. L., 1992, MNRAS, 259, 421
%\bibitem[]{} Cristiani S. et al., 2004, ApJ, 600, L119
\bibitem[]{} Cristiani S. et al., 2003, Proc. of the Meeting
Baryons in Cosmic Structures, Monte Porzio (Italy), ASP. Conf.
Ser., eds. E. Giallongo, G. De Zotti, N. Menci, in press
\bibitem[]{} Croom S. M. et al., 2003, astroph/0403040
\bibitem[]{} de Vaucouleurs G., Olson D. W., 1982, ApJ, 256, 346
\bibitem[]{} Elvis M. et al., 1994, ApJS, 95, 1
\bibitem[]{} Elvis M., Risaliti G., Zamorani G., 2002, ApJ, 565, L75
\bibitem[]{} Faber S. M., Jackson R. E., 1976, ApJ, 204, 668
\bibitem[]{} Faber S. M. et al., 1997, AJ, 114, 1771
\bibitem[]{} Fabian A. C., 2003, in Carnegie Observatories Astrophysics Series,
Vol. 1: Coevolution of Black Holes and Galaxies, ed. L.C. Ho
(Cambridge Univ. Press), in press, astroph/0304122
\bibitem[]{} Fan X. et al., 2001, AJ, 122, 2833
\bibitem[]{} Ferrarese L., 2002, Proceedings of the 2nd KIAS Astrophysics Workshop, Seoul, Korea, astroph/0203047
\bibitem[]{} Ferrarese L., Merritt D., 2000, ApJ, 539, L9
\bibitem[]{} Filippenko A. V., Ho L. C., 2003, ApJ, 588, 13
\bibitem[]{} Fukugita M., Shimasaku K., Ichikawa T., 1995, PASP, 107, 945
\bibitem[]{} Fukugita M., Hogan C.J., Peebles P.J.E., 1998, ApJ, 503, 518
\bibitem[]{} Gebhardt K. et al., 2000, ApJ, 539, L13
\bibitem[]{} Gebhardt K. et al., 2001, AJ, 122, 2469
\bibitem[]{} Gehrels N., 1986, ApJ, 303, 336
\bibitem[]{} Ghez A. M. et al., 2003, ApJ, 586, 127
\bibitem[]{} Gilli R., 2004, in "New Results from Clusters of
Galaxies and Black Holes", Advances in Space Research, Eds. C.
Done, E. M. Puchnarewicz, M. J. Ward. (Amsterdam: Elsevier
Science), in press (astro-ph/0303115)
\bibitem[]{} Gonzalez A. H. et al., 2000, ApJ, 528, 145
\bibitem[]{} Granato G. L., De Zotti G., Silva L., Bressan A., Danese L., 2004, ApJ, 600, 580
\bibitem[]{} Haehnelt M. G., Natarajan P., Rees M. J., 1998, MNRAS, 300,
817
\bibitem[]{} Hasinger G., 2003, in proc.  conf. ``The restless high energy
universe",  E.P.J. van den Heuvel, J.J.M. in 't Zand, and R.A.M.J.
Wijers (eds.), Nucl. Physics B. Suppl. Ser., in press,
astro-ph/0310804
\bibitem[]{} Heckman T. M. et al., astro-ph/0406218
\bibitem[\protect\citeauthoryear{Ho, Filippenko, \&
Sargent}{1997}]{1997ApJ...487..568H} Ho L.C., Filippenko A.V.,
Sargent W.L.W., 1997, ApJ, 487, 568
\bibitem[]{} Hosokawa T., 2002, ApJ, 576, 75
\bibitem[\protect\citeauthoryear{Huchra \&
Burg}{1992}]{1992ApJ...393...90H} Huchra J., Burg R., 1992, ApJ,
393, 90
\bibitem[]{} J\o rgensen I., Franx M., Kjaergaard P., 1995, MNRAS,
276, 1341
\bibitem[]{} Kawakatu N., Umemura M., 2002, MNRAS, 329, 572
\bibitem[\protect\citeauthoryear{King}{2003}]{2003ApJ...596L..27K} King A.,
2003, ApJ, 596, L27
\bibitem[]{} Kochanek C.S., 2001, in Proceedings of The Dark
Universe Meeting at STScI, M. Livio, ed., Cambridge University
Press, astroph/0108160
\bibitem[]{} Kochanek C. S. et al., 2001, ApJ, 560, 566
\bibitem[]{} Kormendy J., 2003, in Carnegie Observatories Astrophysics Series,
Vol. 1: Coevolution of Black Holes and Galaxies, ed. L.C. Ho
(Cambridge Univ. Press), in press, astroph/0306353
\bibitem[]{} Kormendy J., Gebhardt K., 2001, in
20th Texas Symposium on relativistic astrophysics, ed. J.C.
Wheeler, \& H. Martel (AIP Conf. Proc. 586; Melville: AIP), 363
\bibitem[]{} Kormendy J., Richstone D., 1995, ARA\&A, 33, 581
\bibitem[]{} Lynden-Bell D., 1969, Nature, 223, 690
\bibitem[]{} Magorrian J. et al., 1998, AJ, 115, 2285
\bibitem[]{} Marconi A., Hunt L., 2003, ApJL, 589, L21
\bibitem[]{} Marconi A., Risaliti G., Gilli R., Hunt L. K., Maiolino R., Salvati M., 2004, astro-ph/0311619
\bibitem[]{} Martini M., 2004, in Coevolution of Black Holes and Galaxies,
L.C. Ho Ed., Cambridge, p. 170.
\bibitem[]{} McLure R. J., Dunlop. J. S., 2002, MNRAS, 331, 795
\bibitem[]{} McLure R. J., Dunlop. J. S., 2003, MNRAS submitted, astroph/0310267
\bibitem[]{} Merritt D., Ferrarese L., Joseph C. L., 2001, Science, 293, 5532, 1116
\bibitem[]{} Miyoshi M. et al., 1995, Nature, 373, 127
\bibitem[]{} Murali C. et al., 2002, ApJ, 571, 1
\bibitem[]{} Nakamura O. et al., 2002, AJ, 125, 1682
\bibitem[]{} Osmer P. S., 2003, in Carnegie Observatories Astrophysics Series,
Vol. 1: Coevolution of Black Holes and Galaxies, ed. L.C. Ho
(Cambridge Univ. Press), in press, astroph/0304150
\bibitem[]{} Pei Y. C., 1995, ApJ, 438, 623
\bibitem[]{} Salpeter E. E., 1964, ApJ, 140, 796
\bibitem[]{} Salucci P., Szuszkiewicz E., Monaco P., Danese L.,
1999,MNRAS, 307, 637
\bibitem[]{} Sch\"odel R. et al., 2002, Nature, 419, 694
\bibitem[]{} Setti G., Woltjer L., 1989, A\&A, 224, L21
\bibitem[]{} Shankar F., Salucci P., Granato G. L., Danese L., 2003 http://www.exp-astro.phys.ethz.ch/ETH\_Astro\_2003%/posters.html%, in http://www.exp-astro.phys.ethz.ch/ETH\_Astro\_2003/posters.html
\bibitem[]{} Sheth R. K. et al., 2003, ApJ, 594, 225
\bibitem[]{} Shimasaku K., 1993, ApJ, 413, 59
\bibitem[\protect\citeauthoryear{Silk \& Rees}{1998}]{1998A&A...331L...1S}
Silk J., Rees M.~J., 1998, A\&A, 331, L1
\bibitem[]{} Small T. A., Blandford R. D., 1992, MNRAS, 259, 725
\bibitem[]{} Soltan A., 1982, MNRAS, 200, 115
\bibitem[]{} Steed A., Weinberg D. H., 2003, astroph/0311312
\bibitem[]{} Thorne K. S., 1974, ApJ, 191, 507
\bibitem[]{} Tremaine S. et al., 2002, ApJ, 574, 740
\bibitem[]{} Ueda Y., Akiyama M., Ohta K., Miyaji T., 2003, ApJ, 598, 886
\bibitem[]{} van der Marel R. P., 2003, in Carnegie Observatories Astrophysics
Series, Vol. 1: Coevolution of Black Holes and Galaxies, ed. L.C.
Ho (Cambridge Univ. Press), in press, astroph/0302101
\bibitem[]{} Vestergaard M., 2004, ApJ, 601, 676
\bibitem[\protect\citeauthoryear{Vignali, Brandt, \&
Schneider}{2003}]{2003AJ....125..433V} Vignali C., Brandt W.~N.,
Schneider D.~P., 2003, AJ, 125, 433
%\bibitem[]{} Vignali, C., et al. 2003, AJ 125, 2876
\bibitem[]{} Vecchi A. et al., 1999, A\&A, 349, L73-L76
\bibitem[\protect\citeauthoryear{Wandel, Peterson, \&
Malkan}{1999}]{1999ApJ...526..579W} Wandel A., Peterson B.M.,
Malkan M.A., 1999, ApJ, 526, 579
%\bibitem[]{} Wyithe J. S. B., Loeb A., 2003, ApJ, 576, 75
\bibitem[]{} Wyithe J. S. B., Loeb A., 2003, ApJ, 595, 614
\bibitem[]{} Yu Q., Lu Y., 2004, ApJ, 602, 603
\bibitem[]{} Yu Q., Tremaine S., 2002, MNRAS, 335, 965
\bibitem[]{} Zel'dovich Ya. B., Novikov I. D., 1964, Soviet Phys. Dokl.,
158, 811
\end{thebibliography}
\end{document}